\documentstyle[10pt,aas2pp4,psfig, tighten]{article}

\def\sp{\hspace{1.5pt}}
\def\etal{{\it et~al.}}
\def\amin{\ifmmode^{\prime}\else$^{\prime}$\fi}
\def\asec{\ifmmode^{\prime\prime}\else$^{\prime\prime}$\fi}

\def\simgt{\lower.5ex\hbox{$\; \buildrel > \over \sim \;$}}
\def\simlt{\lower.5ex\hbox{$\; \buildrel < \over \sim \;$}}

\newcommand\asca{{\it ASCA}}

\def\sp{\hskip 1.5pt}
\def\psr{\hbox{PSR J1844$-$0258}}
  
\accepted{Mar 28 1998}

\lefthead{Gotthelf \& Vasisht}
\righthead{Discovery of a 7 s X-ray Pulsar}

\begin{document}

\title{\large Discovery of a 7 Second Anomalous X-ray Pulsar in the Distant Milky Way}

\author{E. V. Gotthelf\altaffilmark{1}}
\affil{NASA/Goddard Space Flight Center, Greenbelt, MD 20771}
\authoremail{gotthelf@gsfc.nasa.gov}

\author{G. Vasisht}
\affil{Jet Propulsion Laboratory, California Institute of Technology}
\affil{MS 171-113, Pasadena, CA 91109}
\authoremail{gv@astro.caltech.edu}

\altaffiltext{1}{Universities Space Research Association}

\begin{abstract}

We report the serendipitous discovery of a 7-s X-ray pulsar using data
acquired with the {\it Advanced Satellite for Cosmology and
Astrophysics} (\asca ). The pulsar is detected as an
unresolved source located towards a region of the
Galactic plane ($l,b \simeq 29.5, 0.08$) that coincides with an overdensity of
star-formation tracers. The signal suffers tremendous
foreground absorption, equivalent to $N_H \simeq 10^{23}$ cm$^{-2}$; the
absorption correlates well 
with a line-of-sight that is tangential to the inner spiral
arms and the 4-kpc molecular ring. The pulsar is not associated with
any known supernova remnants or other cataloged objects in that
direction. The near sinusoidal pulse (period $P \simeq 6.9712$) is
modulated at 35\% pulsed amplitude, and the steep
spectrum is characteristic of hot black-body emission with
temperature $kT \sim 0.65$ keV.
We characterize the source as an anomalous X-ray pulsar (AXP).

\end{abstract}

\keywords{pulsars: general --- pulsars: individual (\psr ) --- 
X-rays: general --- supernova remnant}

\section{Introduction}

A canonical young pulsar (period $\sim 100$ ms and stellar dipole
field $\sim 3\times10^{12}$ G) is a rapidly rotating neutron star,
created as the stellar remnant during a Type II (or Ib) supernova
explosion of a massive star. The birthrate of pulsars is known to be
close to $1 - 3$ per century, and it is estimated that there are
about $10^5$ active and $10^8$ defunct neutron stars in the Galaxy
(see Lorimer et al. 1993 and refs. therein).  

In the last few years, there has been growing recognition of a
population of ultra-magnetized neutron stars, or ``magnetars''
(Thompson \& Duncan 1993).  The mostly circumstantial evidence comes
from investigations of the following categories of objects: the soft 
gamma-ray repeaters (Thompson \& Duncan 1995; Frail et al. 1997), long period
pulsars in supernova remnants (Vasisht \& Gotthelf 1997 and
refs. therein), other seemingly isolated, young, long period pulsars
(Thompson \& Duncan 1996) 
nowadays referred to as the anomalous X-ray
pulsars (AXP; van Paradijs et al. 1995), and perhaps their older variants
(Kulkarni \& van Kerkwijk 1998). These objects share some common
properties; they are steady, bright X-ray sources ($L_X \simgt
10^{35}$ erg s$^{-1}$) which show no signs for an accompanying
companion, those with known periods are found to be spinning down, and
all are relatively young ($ \simlt 10^{5}$ yr-old). 

The evolutionary consequences of such large dipole fields are reflected
in the properties listed above.  Most importantly,
large braking torques acting on the star cause it to spin-down rapidly, 
and the magnetic free energy quickly dominates over the
rotation energy, i.e., within several hundred years. For the
above sources, the rotation rates lie between 6 - 12 s, with ages
$\simlt 10^5$ yr (for the SGRs the evidence for periods is indirect,
however, their ages are well constrained due to their association
with supernova remnants).
It is believed that
field decay, which is expected for ultramagnetized neutron stars
(Thompson \& Duncan 1996; also Goldreich \& Reisenegger 1992),
influences the thermal evolution and powers the large X-ray
luminosities observed for the purported magnetars, $L_X \simgt
10^{35}$ erg s$^{-1}$.

If magnetars represent the tail-end of the magnetic field distribution
of neutron stars, then they are bound to be rare. Assume that their
birthrate is 10\% the birthrate of neutron stars
(some justification for this comes from the estimated birthrates of
SGRs; Kulkarni et al. 1994), and that they have active X-ray lifetimes
of $\sim 10^{5}$ yr. These assumptions imply that at present there
are only $\sim 100$ active magnetars in the Galaxy, a conclusion that is borne
out by the observations of the aforementioned objects. The fact that
we observe the five known AXPs through large column densities in the Galaxy,
$N_H \simgt 10^{22}$ cm$^{-2}$, suggests that they are indeed that
rare, and the fact that they are often associated with supernova
remnants or lie near star-formation regions (in spite of the large random
velocities usually attributed to neutron stars; Lyne \& Lorimer 1994)
suggests that they are young. Similarly, only two Galactic SGRs are known,
and it has been suggested that the SGR population census
is nearly complete (Kouveliotou 
1995).

In summary, AXPs have long rotation
periods, hot blackbody-like spectra ($kT \sim$ 0.5 keV) with $10^{35-36}$
erg s$^{-1}$ steady luminosities, and have thus far only been observed
at X-ray wavelengths.
A search for new AXPs in the ASCA database has turned up another
candidate, which we refer to as \psr. 

\section{Observations}

The supernova remnant Kes\sp75 was observed with the
ASCA Observatory (Tanaka et al. 1994) on Dec 10, 1993, as a
Performance and Verification target. Details of the Kes\sp75
observation are already published in Blanton \& Helfand (1996). Herein, we
summarize information that is pertinent to the current analysis.  Data
were acquired by the two gas imaging spectrometers (GIS2 and GIS3) and
collected in the highest time resolution modes ($61 \ \mu \rm{s}$ or
$4.88 \ \rm{ms}$ depending on data acquisition rate).  The GIS offers
$\sim$ 8\% ${(6/E~\hbox{(keV)})}^{1/2}$ spectral resolution over its
$\sim 1-10$ keV energy band-pass.  Each GIS sits at the focus of a
conical foil mirror and the combination results in a spatial
resolution of $1-3$ arcminute (depending on energy) over a $\sim
44^{\prime}$ arcmin diameter active field-of-view.  In this report, we
concentrate exclusively on the GIS data since the target
lies outside the field-of-view of the two other focal plane
instruments (the SISs).

We analyzed data made available from the ASCA public archive, which
were edited to exclude times of high background contamination using
the standard screening criteria.  This rejects time intervals
of South Atlantic Anomaly passages, Earth occultations, bright Earth
limb in the field-of-view, and other periods of high particle
activity.  An effective exposure of $\simeq 4.4 \times 10^4$ s was
obtained with each detector. Event data from both detectors were
co-added and the arrival times of each photon corrected to the solar
system barycenter.

Figure 1 displays a smoothed, flat-fielded, broad-band image of the
Kes 75 observation, centered on the X-ray bright SNR. To the West, at
the edge of the GIS FOV, we find an unresolved point source $18^h 45^m$, 
$-02^{\circ} 58^{\prime}$ (J2000). This source
lies far off-axis, roughly $17^{\prime}$ from the mean GIS optical
axis where its flux suffers strong vignetting; the source is also in a
high background region of the detector.  Although the source has
severe off-axis aberrations and its flux is reduced, the photon
limited detection is quite significant, $\sim 21 \sigma$ (see Gotthelf
\& Kaspi 1998 for a description of the recipe used to infer the
statistical significance).  Due to the off-axis location, and the
reduced GIS spatial-mode resolution ($\sim 1^{\prime} \times
1^{\prime}$ pixels), the position reconstruction is rather crude and
has large uncertainties of order $\sim 3^{\prime}$, larger than those
for a typical on-axis measurement (see Gotthelf 1994).

\section{Timing Results}

Our strategy involved a pulsar search in existing ASCA GIS Galactic plane
fields,
to locate candidate AXPs. The search was optimized to detect
long period ($1-100$ s) pulsars for reasons of computational efficiency. 
A barycentered light curve was generated at each
test location using photons extracted from a $4^{\prime}$ aperture,
binned in 0.5 s time intervals. We then used the Fast Fourier transforms
to find significant periodicities in the data. In the GIS field
containing Kes 75 we detected a single, highly significant peak in the
Fourier power spectral density plot with no overtones, at a period
of $P \sim 7$ s. This detection is centered on the faint source,
localized to a few arcmin on to the edge of the GIS FOV (see Fig
1); it is not observed anywhere else in the image, making it unlikely
to be an instrumental artifact. 

In order to refine the pulse period, we generated a $\chi^2$
periodogram by folding the timeseries.  After extracting 2550 photons,
again using a $4^{\prime}$ radius aperture centered on the source, and
restricting the energy range to $2-8$ keV for optimal sensitivity, we
folded the data into 10 phase bins for each trial period and searched
a range of periods within $\pm 0.5$ s of the expected period,
oversampled with period increments of $0.05 \times {P^2/T}$, were $P$
is the test period and $T$ the observation duration. The periodogram
produced a 14-$\sigma$ detection at the period of $P \simeq 6.9712 \pm
0.0001$ s (see Figure 2). When the substantial background is taken
into account (see \S3), the signal modulation is at 35\% amplitude of
the mean flux in the $4^\prime$ radius aperture. In order to verify
that the pulsed emission indeed arises from the pointlike source,
we generated an image of the pulsar by subtracting the off-pulse flux
from the pulsed emission (figure 1). This reveals just the pulsar, at
the expected location, and no other significant flux in the image
(contours near Kes 75 are artifact of the noisy subtraction in this
region). The pulse profile shows no detectable evidence for energy
dependence.

Given the length of the observation and the source count rate, we were
able to segment the data into $4 \times 20$ ks intervals to search for
variability in the period over the observation time, but none was
found to within the measurement errors of $\sim \pm 4 \times 10^{-4}$
s, implying a period derivative $\dot P < 4.7 \times 10^{-9}$ s
s$^{-1}$ (not terribly constraining). 
In addition, the light curve is consistent with that of a
steady source placed at the edge of the GIS FOV. The low frequency
spectrum shows no obvious pink noise that is typical 
of accretion powered sources. 

\subsection{\bf The Pulsar Spectrum}

We extracted GIS 2 \& 3 source and background spectra from a
$6^{\prime}$ region centered on the pulsar. A sample background was
taken from a nearby off-source region also at the edge of the detector
to match the high local background, mainly due to particle
interaction with the detector walls.  Response matrices were created
following the standard procedure and the spectra and responses of the
two detectors were co-added.  An  absorbed, powerlaw  was first
used to characterize the spectrum. 
The input spectrum was grouped with a minimum of 60
counts per spectral bin, to ensure adequate statistics after background
subtraction.  Both the Galactic plane and instrument backgrounds are
substantial, forcing us to restrict the fits to $1 - 7$
keV, the energy range where the background flux does not dominate the
source.  The power-law produced an acceptable fit (reduced $\chi^2
\simeq 0.9$ for 25 dof), with a steep photon index of $\Gamma =
5.2^{+1.1}_{-1.0}$ and $N_H \simeq 1.1^{+0.3}_{-0.2}\times 10^{23}$
cm$^{-2}$. The steep index is indicative of the tail of an intrinsically
thermal spectrum. An absorbed blackbody also gives a good fit 
(reduced $\chi^2 \simeq 0.8$ for 25 dof) with a temperature fit of $kT \simeq
0.64^{+0.11}_{-0.09}$ keV and a somewhat lower absorption,
$N_H \simeq 6\times 10^{22}$ cm$^{-2}$.
The residuals show no evidence for additional emission components.
We deduce the unabsorbed blackbody flux in the $1-10$ keV band to
be $1.2 \times 10^{-11}$ erg cm$^{-2}$ s$^{-1}$.

We also fit the on-pulse spectrum directly, using the off-pulse data
as a background and thereby derive a slightly harder spectral index of
$\Gamma \simeq 4.7^{+1.8}_{-1.2}$ for the on-pulse emission. This
is independent confirmation that our background subtraction was
successful and gives a reasonable source spectrum, to within the derived 
uncertainties.

\section{The Nature of \psr: A Brief Discussion}

On the basis of its long rotation
period, steady X-ray flux, steep spectral characteristics, and location
in the Galactic plane ($|b| \le 0.5$),
we classify \psr\ as an anomalous X-ray pulsar (see Table 1). The
high foreground absorption suggests that the pulsar is distant, and
its line of sight along the tangent to the Sagittarius-Carina and
Scutum-Crux spiral arms, and the 4-kpc molecular ring justifies its enormous
foreground absorption (see figure 4). Its $N_H$ is roughly twice that
of the nearby remnant Kes 73 for which quoted distances lie between
$10 - 20$ kpc (Blanton \& Helfand 1996). However, the disparity in
$N_H$ does not in itself imply a great dissimilarity in distances.
For instance, at 10 kpc the lines of sight vectors
to these two objects are already separated by
$\sim 100$ pc, the typical sizes and scale heights of dense giant
molecular clouds which are likely to be responsible for most of
the absorbing gas. For the purposes of this article we assume the
distance to be 15 kpc, an estimate likely to be accurate to within
a factor of two.

The steep X-ray spectrum is characteristic of the Wien tail of
a blackbody radiator. Using the best fit blackbody parameters the
isotropic X-ray luminosity is $L_X \simeq 2.5\times 10^{35}d_{15}^2$
erg s$^{-1}$, the distance being 15$d_{15}$ kpc. In effect, the
X-ray pulsations can be ascribed to the viewing of a rotating stellar
hotspot of area $0.15 A_sd_{15}^2$, where $A_s$ is the
area of a neutron star of radius 10 km; this estimate ignores any
relativistic corrections to the inferred area.
Note that the spectrum is unlike that of
any accreting high-mass neutron star binary. Although
such binaries have periods in the range 0.07 - 900 s,
and sometimes go into
low luminosity states with $L_X \sim 10^{35}$ erg s$^{-1}$, they generally
display very hard spectra ($0.8 < \Gamma < 1.5$), and show stochastic
variability on all time-scales (Nagase 1989; Koyama et al. 1989) as is
generally seen in accretion powered sources. We find no evidence for
such variability in our data.

With an AXP classification in hand we can compare the properties of
\psr\ with those of five other members of the AXP family in Table~1.
A few years ago Schwentker (1994) reported weak
5-s pulsations from RX J1838.4$-$0301 which have not yet
been confirmed, Mereghetti et al. (1997)
have argued that this X-ray source might be due to coronal emission
from a late type star. We, therefore, exclude this source from our
list. Although only a future $\dot P$ measurement can help determine the
linear spindown age of \psr\ (an estimator for the age of an isolated neutron star),
its location in the Galactic plane suggests that it is
young, $\tau < 10^5$ yr. We consider it extremely likely that the pulsar is
associated with one of the several star-formation complexes expected to lie
along this line-of-sight (see Fig 4), out to a distance of 20 kpc.
The pulsar lies along a rich region of the
Galaxy; there are 10 supernova remnants, several radio pulsars, and
the $\gamma$-ray source GRO J1838$-$04, all within a $3\times3$ deg$^2$
patch of sky surrounding \psr. However, to the best of our
knowledge no cataloged sources are associated with it.

We now describe other models that address the unique properties of AXPs.
It is often noted in the literature, that the inferred accretion
rate for the pulsars in Table~1 are close to those expected for
accretion powered pulsars with field strengths $B
\sim 10^{11-12}$~G, spinning at their equilibrium periods $P_{eq}$ (see Bhattacharya \& van den Heuvel 1991).
This motivated Mereghetti \& Stella (1995) to suggest that these
pulsars are members of a subclass of low mass X-ray binaries (LMXBs)
in equilibrium rotation, with the stellar magnetic field of order $B_s \sim
10^{11}$ G. In contrast, van Paradijs et al. (1995) argue that these
objects are isolated neutron stars accreting from a fossil disk, while
Ghosh et al. (1997) suggest that AXPs are formed as the result of a
Thorne-$\dot{\rm{Z}}$ytkov phase of a high mass X-ray binary with
strong spherical accretion leading to the soft X-ray spectra with high
foreground absorption. It is worth mentioning that
accretion scenarios would be hard-pressed to
explain the spin-down age of at least one member of
Table 1, the $\sim 2000$ yr of the pulsar in Kes~73
(see Table 1; Gotthelf \& Vasisht 1997). First, it is difficult for
accretion torques to spin-down a pulsar to 12-s in $\sim 10^3$ yr from
initial periods $P_i \simlt 10^2$ ms unless, of course, the pulsar
were born a very slow rotator, which is quite interesting in its own
right.  Secondly, if the pulsar were rotating near its equilibrium
period, as in the Ghosh and Lamb (1979) scenario, the spin-down time
of ${P/ 2\dot P} \sim 3900$ yr is inconsistent with the luminosity implied
accretion rate, $\dot M \simeq 10^{-11}$ M$_\odot$ yr$^{-1}$ (assuming
the pulsar has a standard dipolar field $\simeq 10^{12}$ G); these
usually lie in range $10^4 - 10^5$ yr.

In conclusion, further X-ray observations are required to secure the
classification of the 7-s
pulsar to the growing family of AXPs - by measuring the long term
stability of the X-ray flux, secular trends in the pulse period including
Doppler modulation, the lack of which will
firm up the likelihood against an accreting binary hypothesis.
Indeed, if AXPs are akin to SGRs then they could display sporadic hard
X-ray transients, although such behavior is yet to be observed.
Infrared observations to search for a possible counterpart to \psr\ could
be carried out in spite the somewhat crude localization; we
mention that past optical/IR searches for counterparts have been
unsuccessful.
Furthermore, radio observations to uncover an associated supernova remnant
could be pursued. The high foreground absorption could easily cloak the
soft X-ray emission of an aged, few$\times 10^4$ yr-old, remnant.
We find it remarkable that the period of \psr\ agrees so well with
that of 1E~2259+586, although at present we believe this is sheer
coincidence.

\begin{acknowledgements}
{\noindent \bf Acknowledgments} --- We thank the \asca\ teams for
making these observations possible. This research made use of data
obtained through the HEASARC online service, provided by
NASA/GSFC. E.V.G. is supported by USRA under NASA contract NAS5-32490. We
are grateful to Dr. Erik Leitch for assistance with figure 4.
\end{acknowledgements}

\footnotesize
\begin{deluxetable}{lcccccccc}
\tablecaption{Anomalous X-ray Pulsars and their Properties \vfill
\label{tbl-1}}
\tablehead{
\colhead{Pulsar} & \colhead{SNR} & \colhead{Ref.} & \colhead{P} & \colhead{$P/2\dot P$} & \colhead{$\Gamma$}& \colhead{$kT$} & \colhead{Luminosity$^a$} & \colhead{N$_H$}\nl
      \colhead{} & \colhead{}    & \colhead{}     & \colhead{(s)} & \colhead{(yrs)}      & \colhead{}        & \colhead{(keV)} & \colhead{($\times 10^{35} \rm{ergs~s}^{-1}$)} & ($\times 10^{22}$ cm$^2$)
}
\startdata
\psr\             &   \dots  & \dots & 6.97  &    \dots        & $5$ & 0.64 & $3 d_{15 kpc}^2$  & $10$ \nl
1E 1841$-$045     & Kes 73   & b & 11.76 & $3,400$             & $3.2$ & 0.55 & $3 d_{7 kpc}^2$   & $2$\nl
1E 2259$+$586     & CTB 109  & c & 6.98  & $1.5 \times 10^{5}$ & $4.1$ & 0.41 & $0.8 d_{4 kpc}^2$   & $1$ \nl
4U 0142$+$615     &  \dots   & d & 8.69  & $6\times 10^4$      & $4$ & 0.39 & $10 d_{4 kpc}^2$  & $1$\nl
1E 1048$-$5937    &  \dots   & e & 6.44  & $1\times 10^{4}$    & $3$ & 0.64 & $5 d_{10 kpc}^2$  & $2$ \nl
RX J170849.0$-$400910& \dots & f & 11.00 &    \dots            & $3.5$ &\dots & $10 d_{10 kpc}^2$ & $2$\nl
\enddata
\tablenotetext{a}{All luminosities are corrected for the effect of absorption and are quoted in the $\sim 1-10$ keV energy band}
\tablenotetext{b}{Vasisht \& Gotthelf 1997; $^c$Corbet \etal\ 1995, Iwasawa \etal\ 1992, and refs. therein.;
$^d$White \etal\ 1996, Mereghetti \& Stellar 1995, and refs. therein.;
$^e$Oosterbroek \etal\ 1998, Mereghetti \etal\ 1997, Corbet \& Mihara 1997 and refs. therein.; 
$^f$Sugizaki \etal\ 1997}
\tablenotetext{c} {The 8-s pulsar RXJ 0720.4$-$3125 is not included in this
table as it does not have the properties of a young AXP (Haberl et al. 1997;
Kulkarni \& van Kerkwijk 1998).}
\end{deluxetable}

\onecolumn

\begin{figure}

\centerline{
\psfig{figure=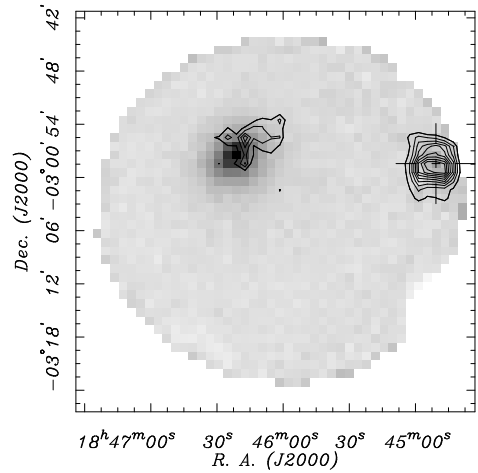,height=3in,angle=270.0,bbllx=25bp,bblly=25bp,bburx=587bp,bbury=550bp,clip=}
}
\caption{
The \asca\ GIS broad-band image (greyscale) of the region containing Kes
75, the central bright source. The location of the pulsar
\psr\ is marked by the cross. The overlayed contour plot shows the
pulsed emission component only. The pulsar is clearly visible towards the West;
the low contours near Kes 75 are not statistically significant. 
The images are scaled by
the square root of the intensity while the contours are spaced in 10\% 
increments.  }
\end{figure}

\begin{figure}
\centerline{
\psfig{file=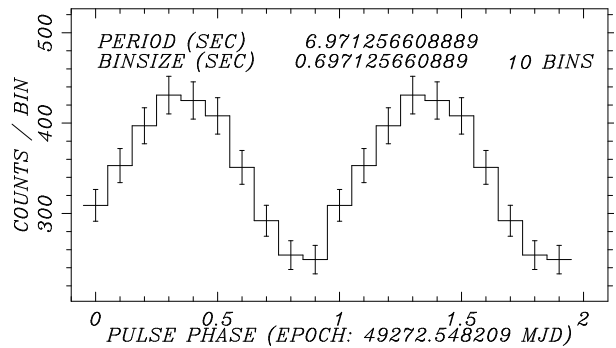,height=3in,angle=270,clip=}
}
\caption{
The folded GIS pulse profile of \psr\ using 10 phase bins and a folding period 
of 6.9712 s. The profile is roughly sinusoidal, with $\sim 35\%$ modulation, after accounting for the background. Two cycles are shown for clarity.
}
\end{figure}

\begin{figure}
\centerline{
\psfig{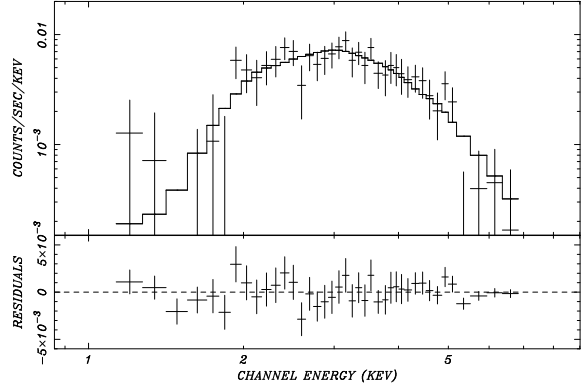}
}
\caption{The background subtracted GIS pulse height
spectrum of the \psr\ emission. GIS 2 and 3 data are summed after gain
corrections. The data are fit with a single powerlaw + foreground 
absorption model, yielding a photon index $\Gamma \simeq 5$ (see
section 3.1 for details).  }

\end{figure}

\begin{figure}
\centerline{
\psfig{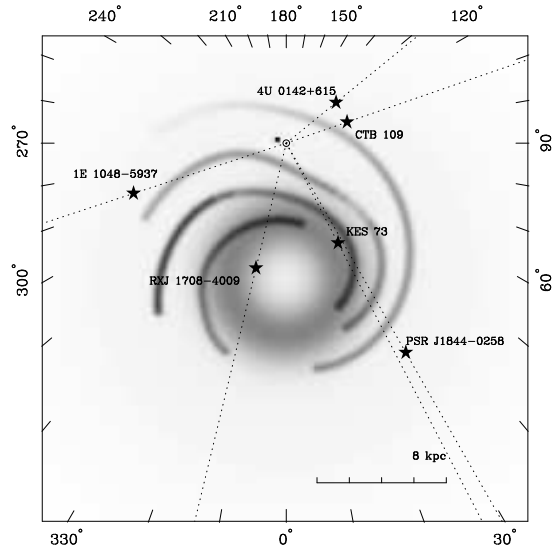}
}
\caption{ This figure illustrates the rough electron density distribution in the
Milky Way tracing out its spiral arms represented in greyscale (Taylor \& Cordes
1993). The
lines-of-sight to sources listed in Table 1 are sketched with the ``stars'' marking
their rough distances and the likely intervening overdense regions may be
seen. The
distribution of AXPs shows that they are rare objects and are likely to be
detectable throughout the Galaxy in 1-10.0 keV X-rays. }

\end{figure}

\end{document}